\documentclass[%
 reprint,
 amsmath,amssymb,
 aps,
]{revtex4-2}

\usepackage{graphicx}

\usepackage{dcolumn}
\usepackage{bm}
\usepackage{braket}
\usepackage{float}
\usepackage{amsmath}
\usepackage{xcolor}

\begin{document}

\preprint{APS/123-QED}

\title{Decoherence-assisted quantum key distribution}


\author{Daniel R. Sabogal$^{1}$}
\altaffiliation[Now at ]{Institute of Photonics and Quantum Sciences (IPAQS), Heriot-Watt University, Edinburgh, UK.}
\author{Daniel F. Urrego$^2$}
\author{Juan Rafael Álvarez$^{4,5}$}
\author{Andrés F. Herrera$^1$}
\author{Juan P. Torres $^{2,3}$}
\author{Alejandra Valencia$^1$}
 \altaffiliation[Also at ]{ dr.sabogal@uniandes.edu.co}
\affiliation{
$^1$ Laboratorio de \'Optica Cu\'antica, Univ. de Los Andes, 4976 Bogot\'a, Colombia.
}

\affiliation{$^2$ICFO – Institut de Ciencies Fotoniques, The Barcelona Institute of Science and Technology, 08860
Castelldefels, Barcelona, Spain.
}
\affiliation{$^3$Department of Signal Theory and Communications, Universitat Politecnica de Catalunya, Barcelona,
Spain.}

\affiliation{$^4$Université Paris-Saclay, CNRS, Centre de Nanosciences et de Nanotechnologies, 91120, Palaiseau, France.}

\affiliation{$^5$Clarendon Laboratory, University of Oxford, Parks Road, Oxford OX1 3PU, United Kingdom.}

\date{\today}

\begin{abstract}
We present a theoretical and experimental study of a controllable decoherence-assisted quantum key distribution scheme. Our method is based on the possibility of introducing controllable decoherence to polarization qubits using the spatial degree of freedom of light. We show that our method reduces the amount of information that an eavesdropper can obtain in the BB84 protocol under the entangling probe attack. We demonstrate experimentally that Alice and Bob can agree on a scheme to that gives low values of the quantum bit error rate, despite the presence of a large amount of decoherence in the transmission channel of the BB84 protocol.

\end{abstract}

\maketitle


\section{\label{sec:level1}Introduction\protect}
Quantum key distribution (QKD) allows two parties to distribute securely a cryptographic key using the principles of quantum mechanics. Different degrees of freedom of light have been used for this purpose, such as polarization~\cite{Poppe:04,Bennett1992}, frequency~\cite{Bloch:07}, continuous variables~\cite{Jouguet2013}, and orbital angular momentum~\cite{Mirhosseini_2015}. The security of some QKD protocols can be proven theoretically. An example of this is the BB84 protocol~\cite{Bennett_2014}, for which it has been demonstrated that an unconditionally secure secret key can be distilled if the quantum bit error rate ($\mathrm{QBER}$) is below $11\%$~\cite{PhysRevLett.85.441, Scarani2009}.

Specific eavesdropper attacks have been considered and analyzed ~\cite{doi:10.1080/00107514.2016.1148333}. For example, the eavesdropper can attack one photon at a time~\cite{PhysRevA.61.052304,PhysRevA.57.2383} by 
ensuring that the photon Alice distributes to Bob interacts, through a unitary transformation, with a probe photon belonging to Eve~\cite{PhysRevA.56.1163}. This method, referred to as the entangling probe attack, has been studied in detail~\cite{PhysRevA.73.012315,PhysRevA.75.042327,PhysRevA.71.042312}. In fact, while the ideal BB84 protocol without any eavesdropper has a $\mathrm{QBER}$=0, under the entangling probe attack, the $\mathrm{QBER}$ increases when the eavesdropper obtains information about the key.

In this paper, we introduce a controllable decoherence-assisted scheme. With this method, it is possible to use the ostensibly detrimental effects of decoherence to increase the security of the BB84 protocol under the entangling probe attack. In
particular, our method allows to reduce the amount of information that an eavesdropper can obtain from attacking the channel set between Alice and Bob. Our method takes advantage of the possibility of introducing decoherence in a controllable way that indeed can be canceled when it is induced appropriately on Alice’s and Bob’s sides. The decoherence is induced with a
dephasing channel implemented using spatial and polarization photonic
degrees of freedom~\cite{Urrego:18}. Specifically, the transverse momentum of light acts as an environment that induces
a tunable dephasing on a quantum system represented by the polarization of light. The coupling between environment and system is controlled by a
parameter that can be adjusted at will.

This paper is organized as follows: In section~\ref{teoria}, we
start by introducing a theoretical model that describes the effect of the controllable decoherence-assisted scheme in the BB84 protocol. We present a brief overview of the entangling probe attack to mathematically demonstrate that the security of the BB84 protocol under such attack presents an improvement when the
controllable decoherence-assisted scheme is used. In section~\ref{experiment}, we present experimental results to demonstrate 
that the specific type of decoherence introduced by a controllable dephasing channel at Bob’s side can cancel dephasing decoherence effects introduced by Alice. We show this effect by inducing the appropriate decoherence in Alice's and Bob’s sides and recovering the $\mathrm{QBER}$ of the BB84 protocol in the absence of an eavesdropper. Finally, in section~\ref{conclusions}, we draw our conclusions.

\begin{figure}
    \centering
    \includegraphics[scale=1]{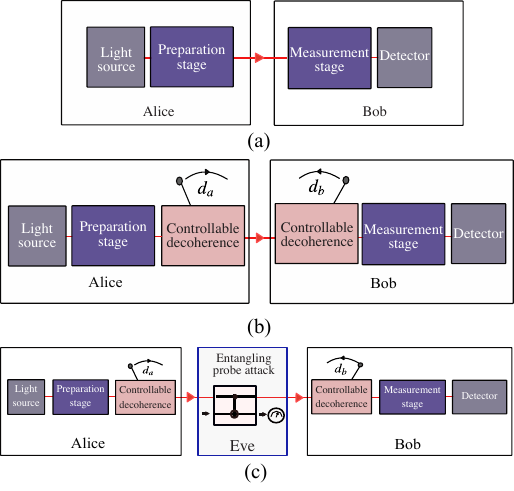}
    \caption{(a) Ideal BB84 protocol: Alice uses a light source that can be randomly prepared in a specific polarization state. Afterwards, Alice sends the light to Bob. On Bob's side, he randomly chooses a basis to measure the state that Alice has sent. (b) BB84 protocol under the controllable decoherence assisted scheme. $d_{a}$ and $d_b$ are parameters that tune the decoherence induced in Alice's and Bob's sides, respectively. (c) Entangling probe attack under the controllable decoherence assisted scheme.}
    \label{fig:schemes}
\end{figure}

\section{Theoretical background}\label{teoria}

In this section, we present the theoretical model behind the controllable decoherence-assisted scheme. We start by considering the ideal situation in which there is no eavesdropper. Then, we present a brief overview of the entangling probe attack and move to consider the effect of our method when the BB84 protocol is under such attack.

\subsection{BB84 protocol and the controllable decoherence-assisted scheme}\label{CDAS}

In the ideal BB84 protocol (Fig.~\ref{fig:schemes}(a)), Alice sends a state that can be randomly prepared in horizontal ($\ket{H}$), vertical ($\ket{V}$), diagonal ($\ket{D}$) or anti-diagonal ($\ket{A}$) polarizations. Bob measures the state by randomly choosing either the $\left\{ \ket{H},\ket{V} \right\}$ or the $\left\{ \ket{D},\ket{A} \right\}$ bases to measure the polarization state that Alice has sent. Alice and Bob then repeat the procedure $n$ times and execute the BB84 protocol to obtain a key of length $\approx n/2 $ ~\cite{Bennett_2014}.

The method we propose relies on the possibility of introducing decoherence in a controllable way on Alice and Bob's sides, as shown in Fig.~\ref{fig:schemes}(b). The induced decoherence is parametrized by $d_a$ and $d_b$ in Alice and Bob sides, respectively. Under the decoherence-assisted scheme, the parties must share one more parameter when executing the BB84 protocol. Analogously to the state preparation and measurement basis stages of the protocol, the values for $d_a$ and $d_b$ have to be selected, and the choice between them is done randomly. This implies that in the BB84 protocol when Alice and Bob use the controllable decoherence-assisted scheme, one step is added: The reconciliation stage consists on constructing a key from the bits that have the same values of $d_a$ and $d_b$ and the same polarization basis for preparation and measurement.

For the implementation of the decoherence-assisted scheme, the state that Alice initially prepares in the BB84 protocol, $\ket{\Psi}_{\cal{A}}$, must now contain also the spatial degree of freedom, i.e., 
 \begin{equation}
    \ket{\Psi}_{\cal{A}}= \int dy f(y)\Big(\alpha \ket{H,y}_{\cal{A}} + \beta\ket{V,y}_{\cal{A}}\Big),
\end{equation}
with $\alpha$ and $\beta$ complex numbers satisfying $|\alpha|^{2}+ |\beta^{2}| = 1$. $f(y)$ is a normalized function describing the spatial mode of the light. The decoherence can be induced, for example, by using two controllable dephasing channels such as the one described in reference~\cite{Urrego:18}. The dephasing channel, represented by the unitary operator $\hat{U}(d)$, couples the spatial mode, $f(y)$, with the polarization state. As a result of the coupling, the spatial mode gets shifted by a distance $\pm d$. Mathematically, 

\begin{subequations}
\label{eq:Uposition}
\begin{eqnarray}
\hat{U} (d)\int dy~f(y) \ket{H,y} = \int dy~f(y-d) \ket{H,y}
\label{subeq:1position}
\end{eqnarray}
\begin{eqnarray}
\hat{U} (d)\int dy~f(y) \ket{V,y} = \int dy~f(y+d) \ket{V,y}.
\label{subeq:2position}
\end{eqnarray}
\end{subequations}
After transmission, Bob's photon in the quantum state
\begin{equation}
   \hat{\rho}^{(\cal{B})}=\hat{U}(-d_b) \hat{U}(d_a) \hat{\rho}^{(\cal{A})}\hat{U}^{\dag}(d_a) \hat{U}^{\dag}(-d_b),
   \label{eq:operators}
\end{equation}
where $\hat{\rho}^{(\cal{A})}=\ket{\Psi}_{\cal{A}}\bra{\Psi}_{\cal{A}}$ and
the parameter $d$ of the unitary operators corresponds to $d_a$ ($-d_b$) for Alice's (Bob's) dephasing channel. 

The degree of similarity between Alice's and Bob's keys can be monitored using the quantum bit error rate ($\mathrm{QBER}$). This quantity is defined as the probability that a bit on Bob's key is different from the corresponding bit on Alice's key. For the BB84 protocol, 
\begin{equation}
    \mathrm{QBER} = p_{H}^{(\cal{A})}q_{H}^{(\cal{B})} + p_{V}^{(\cal{A})}q_{V}^{(\cal{B})} + p_{D}^{(\cal{A})}q_{D}^{(\cal{B})} + p_{A}^{(\cal{A})}q_{A}^{(\cal{B})}, 
    \label{eq:QBER}
\end{equation}
where $p_{i}^{(\cal{A})}$ is the probability that Alice prepares the state $\ket{i}$. Similarly, $q_{i}^{(\cal{B})}=1-p_{i}^{(\cal{B})}$, where $p_{i}^{(\cal{B})}$ is the probability that Bob measures the same state that Alice had originally prepared. 

Operationally, Alice and Bob can obtain the $\mathrm{QBER}$ by calculating the ratio between the number of unequal bits and the number of total bits in a random portion of the key. Under ideal conditions, $\mathrm{QBER}=0$ after the reconciliation stage.

To calculate the $\mathrm{QBER}$ after applying the decoherence-assisted scheme, it is necessary to calculate the probabilities in Eq.~\ref{eq:QBER}. This requires calculating the partial trace of $ \hat{\rho}^{(\cal{B})}$ over the spatial variables to obtain $\hat{\rho}^{(\cal{B})}_P$, the polarization density matrix for Bob's state. By doing so, 
\begin{equation}
\label{eq:whole2}
\hat{\rho}^{(\cal{B})}_P=\left( \begin{array}{cc}
    |\alpha|^2 &  \alpha^* \beta \gamma_c^*\\
    \alpha \beta^* \gamma_c &  |\beta|^2
\end{array}\right),
\end{equation}
where $\gamma_c=\int_{-\infty}^{\infty} dy f^*(y+d_a-d_b)f(y-d_a+d_b)$ satisfies $0\le|\gamma_c|\le1$. For $d_a=d_b$, $|\gamma_c|=1$ indicating that the decoherence has been compensated for and the polarization state is pure. Conversely, when $d_a$ and $d_b$ are different and larger than the beam width of $f(y)$, $\gamma_c=0$ and the polarization state is maximally mixed.

From Eq.~(\ref{eq:whole2}), one sees that if Alice sends horizontal ($|\beta|^2=0$) or vertical ($|\alpha|^2=0$) polarization states, the decoherence does not affect the states and 
\begin{subequations}
\label{eq:qhbqvb}
\begin{eqnarray}
q_{H}^{(\cal{B})}=1-p_{H}^{(\cal{B})}=1-\bra{H} \hat{\rho}^{(\cal{B})}_P \ket{H}=0,
\label{subeq:qhb}
\end{eqnarray}\begin{eqnarray}
q_{V}^{(\cal{B})}=1-p_{V}^{(\cal{B})}=1-\bra{V} \hat{\rho}^{(\cal{B})}_P \ket{V}=0.
\label{subeq:2position}
\end{eqnarray}
\end{subequations}
On the other hand, when Alice sends a diagonal or anti-diagonal polarization state ($|\alpha|^2=|\beta|^2=1/2$), 
\begin{eqnarray}
q_{A}^{(\cal{B})}&=&1-\bra{A} \hat{\rho}^{(\cal{B})}_P \ket{A}=q_{D}^{(\cal{B})}=1-\bra{D} \hat{\rho}^{(\cal{B})}_P \ket{D}\nonumber \\
&=&\frac{1}{2}-\frac{1}{2}\operatorname{Re}(\gamma_c).
\label{eq:qabqdb}
\end{eqnarray}
Equation~(\ref{eq:qabqdb}) reveals that, under the controllable decoherence-assisted scheme, it is possible to cancel decoherence when $d_a$ = $d_b$, i.e., $\operatorname{Re}(\gamma_c)=1$, obtaining $q_{A}^{(\cal{B})}=q_{D}^{(\cal{B})}=0$. This implies that under these conditions, Bob is capable of retrieving the same polarization sent by Alice. Conversely, for $\operatorname{Re}(\gamma_c)=0$, one gets $q_{A}^{(\cal{B})}=q_{D}^{(\cal{B})}=1/2$, indicating that Bob is not able to identify which polarization state ($\ket{D}$ or $\ket{A}$) was sent by Alice. 

All the polarization states are prepared by Alice with the same probability, $p_{H}^{(\cal{A})}= p_{V}^{(\cal{A})} = p_{D}^{(\cal{A})}= p_{A}^{(\cal{A})}= 1/4$. Therefore, using the probabilities in Eq.~(\ref{eq:qabqdb}), the $\mathrm{QBER}$ given by Eq.~\eqref{eq:QBER} becomes 
\begin{equation}
\mathrm{QBER}(d_{a},d_{b})=\frac{1}{4}[1-\operatorname{Re}(\gamma_c)],
\label{eq:QBER_scheme}
\end{equation} 
revealing that, when Bob receives a maximally mixed polarization state, $\gamma_c=0$ and $\mathrm{QBER}=1/4$. In sharp contrast, when $d_a=d_b$ the decoherence can be compensated. This feature makes the decoherence-assisted scheme powerful since it allows to recover $\mathrm{QBER}=0$ in the presence of controllable decoherence. 

\subsection{Entangling probe attack}
In the entangling probe attack \cite{PhysRevA.73.012315}, Eve intercepts the state that Alice sends to Bob and entangles it with her probe qubit using a C-NOT gate. The intercepted qubit is sent to Bob, and Eve keeps the probe qubit. The entanglement that the C-NOT gate generates between the two qubits allows Eve to obtain information about the polarization state of the intercepted qubit, and is the tool used by Eve to guess the key that Alice and Bob share.

The input quantum state of the probe qubit belonging to Eve can be written as
\begin{equation}
    \ket{T_{in}}_{\cal{E}}= \sqrt{1-S^2}\ket{+}_{\cal{E}}+S\ket{-}_{\cal{E}},
\end{equation}
where the parameter $S$ takes a value in the range between $0$ and $1$, $\ket{\pm}_{\cal{E}}=[\ket{0}_{\cal{E}} \pm\ket{1}_{\cal{E}}]/\sqrt{2}$ with $\ket{0}_{\cal{E}}=\cos(\pi/8)\,\ket{H}_{\cal{E}} + \sin (\pi/8)\,\ket{V}_{\cal{E}}$ and $\ket{1}_{\cal{E}}=-\sin(\pi/8)\,\ket{H}_{\cal{E}} +\cos (\pi/8)\,\ket{V}_{\cal{E}}$. The transformations after the C-NOT operation are
\begin{subequations}
\label{eq: CNOT_transform}
\begin{equation}
\ket{H}_{\cal{A}}\ket{T_{in}}_{\cal{E}}\xrightarrow[]{}\ket{H}_{\cal{B}}\ket{T_{+}}_{\cal{E}}+\ket{V}_{\cal{B}}\ket{T_{\Tilde{e}}}_{\cal{E}}
\end{equation}
\begin{equation}
    \ket{V}_{\cal{A}}\ket{T_{in}}_{\cal{E}}\xrightarrow[]{}\ket{V}_{\cal{B}}\ket{T_{-}}_{\cal{E}}+\ket{H}_{\cal{B}}\ket{T_{\Tilde{e}}}_{\cal{E}}
\end{equation}
\begin{equation}
    \ket{D}_{\cal{A}}\ket{T_{in}}_{\cal{E}}\xrightarrow[]{}\ket{D}_{\cal{B}}\ket{T_{+}}_{\cal{E}}+\ket{A}_{\cal{B}}\ket{T_{\Tilde{e}}}_{\cal{E}}
\end{equation}
\begin{equation}
    \ket{A}_{\cal{A}}\ket{T_{in}}_{\cal{E}}\xrightarrow[]{}\ket{A}_{\cal{B}}\ket{T_{-}}_{\cal{E}}+\ket{D}_{\cal{B}}\ket{T_{\Tilde{e}}}_{\cal{E}},
\end{equation}
\end{subequations}
where $\ket{T_{\pm}}_{\cal{E}}= \sqrt{1-S^2}\ket{+}_{\cal{E}} \pm S/\sqrt{2}\ket{-}_{\cal{E}}$ and $\ket{T_{\Tilde{e}}}_{\cal{E}}=S/\sqrt{2}\ket{-}$.

For Eve to recover the information about the key, she only needs to discriminate between the two states $\ket{T_+}_{\cal{E}
}$ and $\ket{T_{-}}_{\cal{E}}$. However, the entangling-probe attack also provides a penalty for Eve: Her attempt to obtain more information about the key is detrimental to the generation of a valid key between Alice and Bob. Indeed, for the entangling probe attack, the $\mathrm{QBER}$ is estimated to be $S^{2}/{2}$.

To quantify the amount of information learned by Eve, one can use the R\'enyi information~\cite{PhysRevA.57.2383} that can be written as
\begin{align}\label{eq: Renyi_def2} 
  I_{R} &= -\log_{2}\Bigg[\sum_{b =0}^{1} P^{2}(b) \Bigg] \\ \nonumber
  & \hspace{1cm}+ \sum_{e = 0}^{1} P(e) \log_{2} \Bigg[\sum_{b =0}^{1} P^{2}(b|e)\Bigg], 
\end{align}
where $b=\{0,1\}$ and $e=\{0,1\}$ denote the bit values that Bob and Eve obtain during the protocol, respectively. $P(b)$ ($P(e)$) is the prior probability that Bob (Eve) obtains the bit value $b~(e)$, and $P(b|e)$ is the conditional probability that Bob gets a bit with a value $b$ given that Eve has a bit with value $e$.

It has been demonstrated that the R\'enyi information in the entangling-probe attack becomes~\cite{PhysRevA.73.012315,PhysRevA.75.042327}
\begin{equation}
    I_R=\log_2\Bigg[1 + \frac{2S^2(1-S^2)}{(1-S^2/2)^2} \Bigg]
\end{equation}
and that Eve can obtain up to half of the maximum amount of R\'enyi information for $\mathrm{QBER}\leq11\%$, where Alice and Bob meet the security threshold. This result is independent of the choice between $H$-$V$ or $D$-$A$ basis.

\newpage
\subsection{Entangling probe attack and the controllable decoherence-assisted scheme}

Figure~\ref{fig:schemes}(c) illustrates the use of the controllable decoherence-assisted scheme when the communication channel between Alice and Bob is under the entangling probe attack. To calculate the information available to Eve, we calculate the density matrix $\hat{\rho}_j^{(\cal{E})}$ with $j={H, V,D,A}$ by proceeding as follows: First, we dephase Alice's qubit by applying $\hat{U}(d)$ to the input state $\ket{\Psi}_{\cal{A}}$. Second, following Eqs.~\eqref{eq: CNOT_transform}, we apply the C-NOT gate of the entangling probe attack using the dephased qubit as control qubit, and the photon prepared by Eve in $\ket{T_{in}}$ as target qubit. Third, we apply the dephasing channel in Bob's side using $\hat{U}(-d).$ After these steps, the transformations that the four possible input polarization states undergo result in a shared state between Eve and Bob, $\ket{\psi_j}_{\cal{B},\cal{E}}$, and have the form \begin{widetext}
\begin{minipage}{0.9\linewidth}

\begin{subequations}
\begin{align}
\int dyf(y)\ket{H,y}_{{\cal {A}}}\ket{T_{in}}_{{\cal {E}}}\xrightarrow{}\ket{\psi_{H}}_{{\cal B},{\cal E}}=\int dyf(y)\ket{H,y}_{{\cal {B}}}\ket{T_{-}}_{{\cal {E}}}+\int dyf(y-2d)\ket{V,y}_{{\cal {B}}}\ket{T_{\tilde{e}}}_{{\cal {E}}},
\end{align}
\begin{align}
\int dyf(y)\ket{V,y}_{{\cal {A}}}\ket{T_{in}}_{{\cal {E}}}\xrightarrow{}\ket{\psi_{V}}_{{\cal B},{\cal E}}=\int dyf(y)\ket{V,y}_{{\cal {B}}}\ket{T_{+}}_{{\cal {E}}}+\int dyf(y+2d)\ket{H,y}_{{\cal {B}}}\ket{T_{\tilde{e}}}_{{\cal {E}}},
\end{align}
\begin{multline}
\int dyf(y)\ket{D,y}_{{\cal {A}}}\ket{T_{in}}_{{\cal {E}}}\xrightarrow{}\ket{\psi_{D}}_{{\cal B},{\cal E}}=\frac{1}{2}\int dy\Bigg[f(y)\Big(\ket{T_{-}}_{{\cal {E}}}+\ket{T_{+}}_{{\cal {E}}}\Big)+\Big(f(y-2d)+f(y+2d)\Big)\ket{T_{\tilde{e}}}_{{\cal {E}}}\Bigg]\ket{D,y}_{{\cal {B}}}\\
+\frac{1}{2}\int dy\Bigg[f(y)\Big(\ket{T_{-}}_{{\cal {E}}}-\ket{T_{+}}_{{\cal {E}}}\Big)+\Big(f(y+2d)-f(y-2d)\Big)\ket{T_{\tilde{e}}}_{{\cal {E}}}\Bigg]\ket{A,y}_{{\cal {B}}},
\end{multline}
\begin{multline}
\int dyf(y)\ket{D,y}_{{\cal {A}}}\ket{T_{in}}_{{\cal {E}}}\xrightarrow{}\ket{\psi_{A}}_{{\cal B},{\cal E}}=\frac{1}{2}\int dy\Bigg[f(y)\Big(\ket{T_{-}}_{{\cal {E}}}+\ket{T_{+}}_{{\cal {E}}}\Big)-\Big(f(y-2d)+f(y+2d)\Big)\ket{T_{\tilde{e}}}_{{\cal {E}}}\Bigg]\ket{A,y}_{{\cal {B}}}\\
+\frac{1}{2}\int dy\Bigg[f(y)\Big(\ket{T_{-}}_{{\cal {E}}}-\ket{T_{+}}_{{\cal {E}}}\Big)+\Big(f(y-2d)-f(y+2d)\Big)\ket{T_{\tilde{e}}}_{{\cal {E}}}\Bigg]\ket{D,y}_{{\cal {B}}}.
\end{multline}
    \label{eq: CNOT_transformation_dec}
\end{subequations}

\end{minipage}
\end{widetext}Eve's density matrix is obtained (cf. Appendix 1) considering only the error-free part of the state shared between Bob and Eve, called $\ket{\tilde{\psi}_{H}}_{{\cal B},{\cal E}}$, calculated by projecting the state $\ket{\psi_j}_{\cal{B},\cal{E}}$ in Bob's polarization, $\ket{j}_{\cal{B}}$, obtaining $\ket{\tilde{\psi}_j}_{{\cal B},{\cal E}} = \bra j_{{\cal B}}\ket{\psi_{j}}_{{\cal B},{\cal E}}$. This polarization has to match the one prepared by Alice. Since Bob carries out his measurements using a bucket detector (erasing spatial information), the density matrix of Eve is\begin{equation}
\hat{\rho}_{j}^{\left({\cal E}\right)}=\text{Tr}{}_{\text{env}}\big\{\ket{\tilde{\psi}_{j}}_{{\cal B},{\cal E}}\bra{\tilde{\psi}_{j}}_{{\cal B},{\cal E}}\big\}.
\end{equation}

A close examination of Eq.~\ref{eq: CNOT_transformation_dec} reveals that, unlike the case without the controllable decoherence-assisted scheme in Eq.~(\ref{eq: CNOT_transform}), due to the asymmetry between the $H$-$V$ and $D$-$A$ bases, Eve now needs to discriminate between four states to get information about Bob's key: $\{\hat{\rho}_{H}^{({\cal E})},\hat{\rho}_{V}^{({\cal E})}\}$ ($\ensuremath{\{\hat{\rho}_{D}^{({\cal E})},\hat{\rho}_{A}^{({\cal E})}\}}$) when Alice prepares in the $H$-$V$ ($D$-$A$) basis. According to Eq.~\ref{eq: Renyi_def2}, to obtain the Rényi information it is necessary to calculate $P(b|e)$. This conditional probability depends on the basis used for Alice and Bob in the preparation and measuring stages. Therefore, one has to calculate two conditional probabilities: $P_{HV}(e|b)$ for the $\left\{ \ket{H},\ket{V} \right\}$ basis and $P_{DA}(e|b)$, for the $\left\{ \ket{D},\ket{A} \right\}$ basis. These two conditional probabilities can be calculated using the minimum error probability of discriminating two mixed states; which is given by the Helmstrom bound \cite{HELSTROM1967254}:
\begin{eqnarray}
    & & P_{HV}(e|b)=\frac{1}{2}\Big[1-D(\hat{\rho}_H^{(\cal{E})}, \hat{\rho}_V^{(\cal{E})})\Big], \label{eq: probconda1} \\
    & & P_{DA}(e|b)=\frac{1}{2}\Big[1-D(\hat{\rho}_D^{(\cal{E})}, \hat{\rho}_A^{(\cal{E})})\Big],
    \label{eq: probconda2}
\end{eqnarray}
where $D(\hat{\rho}_1, \hat{\rho}_2)$ is the trace distance between the density matrices $\hat{\rho}_1$ and $\hat{\rho}_2$.


An explicit calculation of Eqs.~\ref{eq: probconda1} and \ref{eq: probconda2} (see Appendix 2) can be made by considering a realistic spatial distribution, assuming $f(y)=(2/\pi \text{w}^{2})^{1/4}\exp(-y^{2}/\text{w}^{2})$, i.e., a Gaussian shape with a beam width $\text{w}$. In this case, 
\begin{equation}
    P_{HV}(e|b)=\frac{S^2-2+2S\sqrt{2-2S^2}}{2(S^2-2)}\label{eq: probcond1} 
\end{equation} and \begin{equation}
    P_{DA}(e|b)=\frac{4-4S\sqrt{2-2S^2}\gamma_0 + S^2(\gamma_0^4-3)}{8+2S^2(\gamma_0^4-3)}, 
    \label{eq: probcond2}
\end{equation}
where $\gamma_0=\exp(-2 d^2/\text{w}^2)$ and $d=d_a-d_b$. $\gamma_0$ quantifies the amount of decoherence introduced. When $d\gg \text{w}$, $\gamma_0 \xrightarrow[]{}0$ and the system undergoes complete decoherence. When $d=0$, $\gamma_0=1$ and the system remains in a pure state. 

In the BB84 protocol, Alice randomly switches between polarization bases, so the total R\'enyi information $I_{R}= (I_{R}^{HV}+ I_{R}^{DA})/2$. Subtituting Eq.~(\ref{eq: probcond1}) and Eq.~(\ref{eq: probcond2}) in Eq.~(\ref{eq: Renyi_def2}), the R\'enyi information in the $H$-$V$ basis and in the $D$-$A$ basis become, respectively \begin{equation}
    I^{HV}_R=\log_2\Bigg[1 + \frac{2S^2(1-S^2)}{(1-S^2/2)^2} \Bigg]\label{eq:IHV}
\end{equation} and
\begin{equation}
    I^{DA}_R= \log_2\Bigg[1+\frac{32S^2(S^2-1)\gamma_0^2}{(4+S^2(\gamma_0^4-3))^2}\Bigg]\label{eq:IAD}.
\end{equation}

On the one hand, Eq.~\ref{eq:IHV} is in agreement with the R\'enyi information reported in ~\cite{PhysRevA.73.012315,PhysRevA.75.042327} and does not depend on $\gamma_0$ since the dephasing channel being used does not induce decoherence in the $HV$ basis. On the other hand, Eq.~\ref{eq:IAD} shows that the R\'enyi information depends on the parameter $\gamma_0$, indicating that the use of the controllable decoherence-assisted scheme has implications on the amount of information that Eve can obtain in the $DA$ basis.
Two limiting cases arise: When $\gamma_0 \xrightarrow[]{}0$, the information obtained by Eve in the $DA$ basis is zero, and therefore, Eve can only get information from the key when Alice and Bob use the $HV$ basis. When $\gamma_0=1$, $I^{DA}_R=I^{HV}_R$. As a consequence, $\gamma_0=1$ is the scenario that permits Eve to get more information about the secret key.

The effect of our scheme in the BB84 protocol under the entangling probe attack is highlighted by considering the relationship between the R\'enyi information and the $\mathrm{QBER}$ generated by Eve's presence. Under our scheme, the symmetry between the $HV$ and $DA$ bases is lost, implying that the $\mathrm{QBER}$ in each basis is different. In a calculation analogous to the one in Section \ref{CDAS}, the QBER produced by Eve's presence in each basis is
\begin{equation}
    \mathrm{QBER}_{HV}=S^2/2 
    \label{eq:QBERHV}
\end{equation} and \begin{equation}
    \mathrm{QBER}_{DA}=\frac{1}{4}\Big(3-\gamma_0^4 \Big)S^2. 
    \label{eq:QBERDA}
\end{equation} 

\begin{figure}[t!]
    \centering
    \includegraphics[scale=0.42]{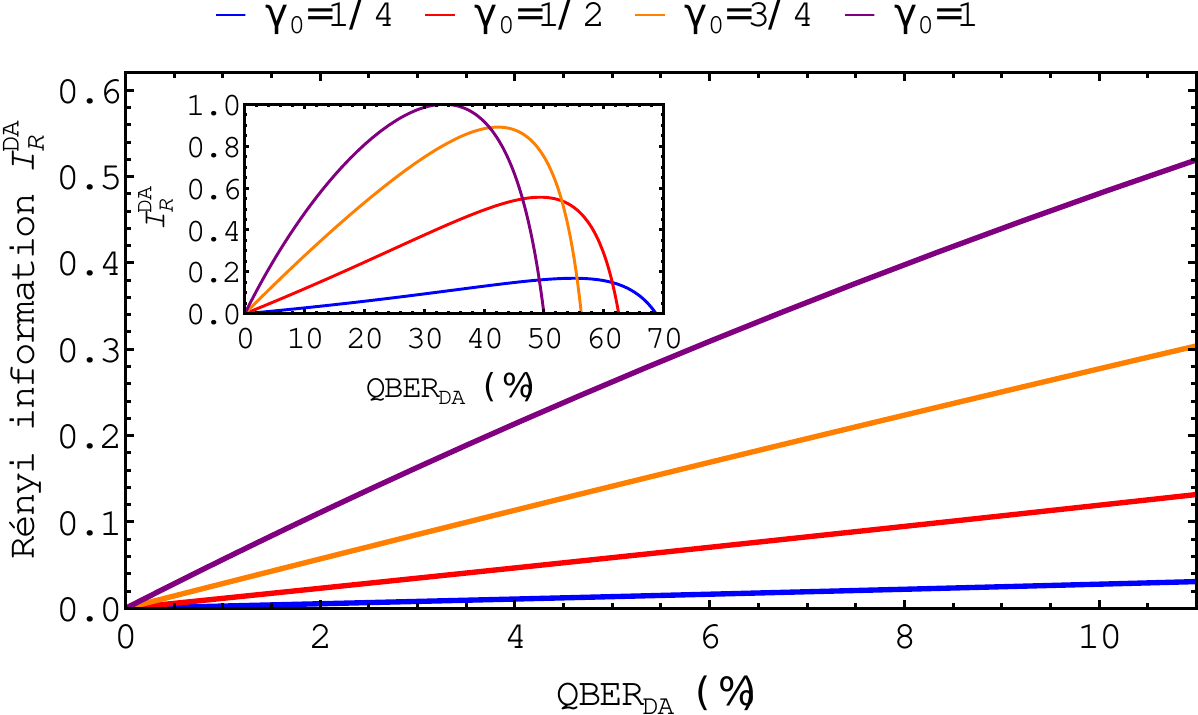}
    \caption{R\'enyi information $I_{R}^{DA}$ as a function of $\mathrm{QBER}_{DA}$ when Eve uses the entangling probe attack with the controllable decoherence-assisted scheme. The inset shows the same function for a wider range of values of $\mathrm{QBER}_{DA}$.}
    \label{fig:Renyi_Advantage_DA}
\end{figure}

Following Eq.~\ref{eq:QBERDA}, it is possible to obtain $S$ as a function of $\mathrm{QBER}_{DA}$, establishing a relation between $I_{R}^{DA}$ and $\mathrm{QBER}_{DA}$. This relation is shown in Fig.~\ref{fig:Renyi_Advantage_DA}(a) in the range where the $\mathrm{QBER}_{DA}<11\%$. The inset shows $I_{R}^{DA}$ for a larger range of $\mathrm{QBER}_{DA}$. From Fig.~\ref{fig:Renyi_Advantage_DA}(a), it is possible to see that for a fixed value of $\mathrm{QBER}_{DA}$, the information that Eve learns decreases as $\gamma_0$ increases. Analogously, Eqs.~\ref{eq:QBERHV} and \ref{eq:QBERDA} enable the calculation of a total $\mathrm{QBER}$ defined as the average between $\mathrm{QBER}_{HV}$ and $\mathrm{QBER}_{DA}$. By doing so, it is possible to establish a relation between $I_R$ and the total $\mathrm{QBER}$, as shown in Fig.~\ref{fig:Renyi_Advantage_DA}(b). When $\gamma_{0}\to0$, the total $I_R$ is saturated at 0.5 and Eve can only obtain information from the $HV$ basis. The reduction of the maximum R\'enyi information available in the controllable decoherence-assisted scheme constitutes an improvement of the security of the BB84 protocol under the entangling probe attack.

\begin{figure}[t]
    \centering
    \includegraphics[scale=0.42]{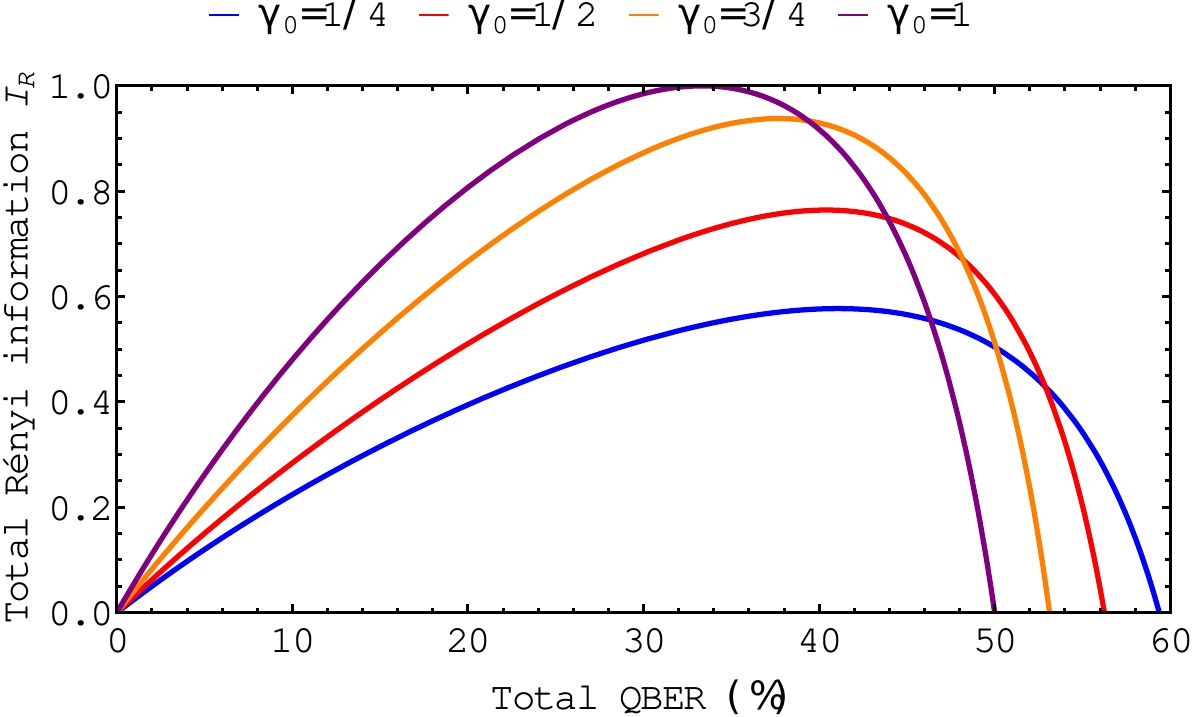}
    \caption{Total R\'enyi information $I_{R}$ as a function of the total $\mathrm{QBER}$ when Eve uses the entangling probe attack with the controllable decoherence-assisted scheme. For values of $\gamma_0<1$, the total Rényi information available to Eve is lower due to the effect of the controllable decoherence assisted scheme. }
    \label{fig:Renyi_Advantage_Total}
\end{figure}

\section{Experiment}\label{experiment}


\begin{figure*}
\includegraphics[width=\textwidth]{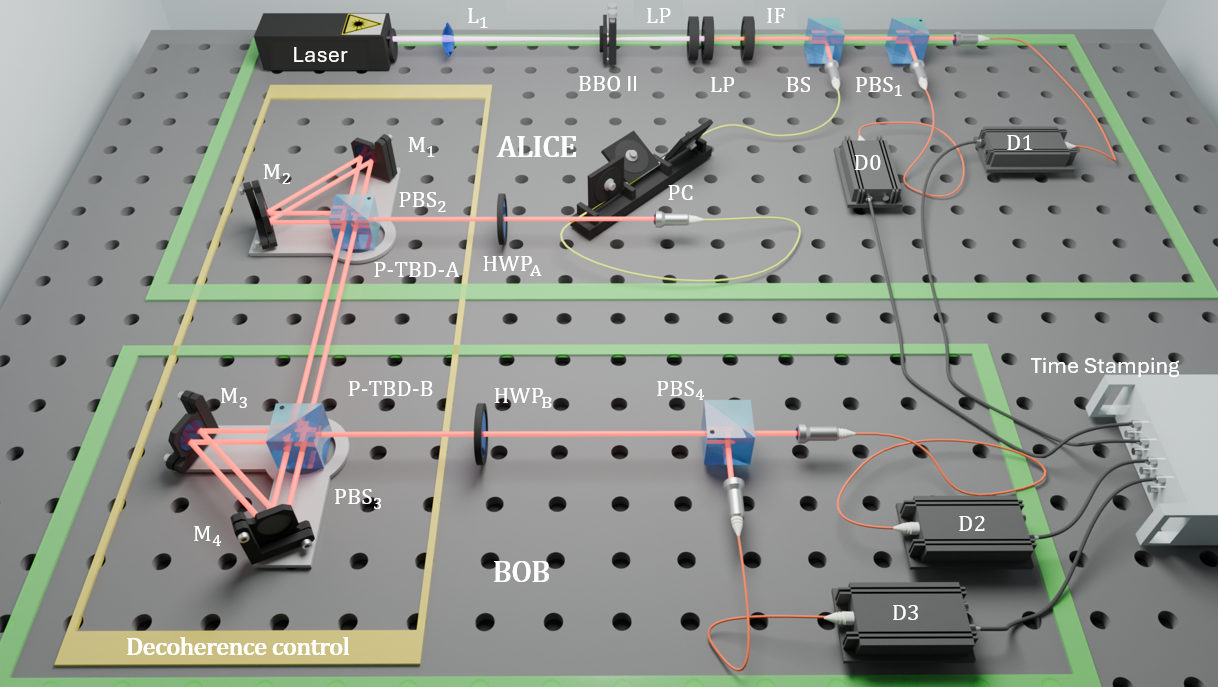}
\caption{\indent \label{fig:setup} Experimental setup. A 407~nm laser pump is used to create photon pairs in a collinear 4 mm BBO type II crystal. The pump's polarization state is adjusted using a half-wave plate (HWP) and the pump is focused into the crystal using a lens ($L_1$, with focal length $f= 100\ \mathrm{mm}$). After the crystal, the pump is removed using two long pass (LP) filters with cut-off wavelengths of 750~nm and an interference filter (IF) of 810$ \pm10$~nm. The idler photon is transmitted in the BS and it is used to herald the presence of its pair taking into account polarization using $\mathrm{PBS}_1$. The signal photon is reflected in the PBS and then collected into the SMF to obtain a Gaussian mode. The polarization controller (PC) is used to correct the polarization state of the idler photon. After the PC, the spatial mode has a beam waist of $\text{w}=0.8$~mm that ensures that the light is collimated throughout its whole optical path during the experiment. A rotating half-wave plate ($\mathrm{HWP}_A$) constitutes the preparation stage. Subsequently, decoherence is induced and reversed by inserting P-TBD-A and P-TBD-B, respectively. On Bob's side, $\mathrm{HWP}_B$ and $\mathrm{PBS}_4$ constitute the measurement stage. All the detections are recorded by a time stamping device. } 
\end{figure*}

Two proof-of-principle experiments were performed to demonstrate that the controllable decoherence-assisted scheme allows to reverse the decoherence effects introduced in Alice's side, thus recovering a low $\mathrm{QBER}$ in the BB84 protocol. In the first experiment, the BB84 protocol was implemented using a heralded single-photon source. In the second one, the controllable decoherence-assisted scheme was introduced. Both experiments are based on the setup shown in Fig.~\ref{fig:setup}. For the first experiment, the decoherence control was not used, which is equivalent to setting $d=0$. Operationally, this was performed by removing the polarizing beam splitters $\mathrm{PBS}_2$ and $\mathrm{PBS}_3$. For the second experiment, $\mathrm{PBS}_2$ and $\mathrm{PBS}_3$ are set to introduce controlled decoherence.

\subsection{Experiment 1: BB84 protocol}
The photons sent by Alice are produced by a heralded single-photon (HSP) source based on spontaneous parametric down conversion (SPDC). The SPDC photon pairs traverse a beam splitter (BS) and the reflected photon is used as the heralded photon to be sent to Bob. This heralded photon passes through a single mode fiber with a polarization controller (PC) to define the photon polarization state and to obtain a Gaussian spatial distribution.

The recognition of the HSP is done by means of the temporal second order correlation function, $G^{2}(\tau)$. Specifically, recognizing pairs of photons that are within a window of width $2 \sigma$, centered at $\tau_{0}$, the maximum of $G^{2}(\tau)$. The value of $\sigma$ is chosen by approximating the measured $G^{2}(\tau)$ to the standard deviation of a Gaussian function. The Gaussian distribution is assumed since the response time of the detectors dominates the shape of the $G^{2}(\tau)$ for SPDC.

In order to make our measurements more efficient, the HSP source was based on Type II SPDC followed by a BS. In this way, the heralding photon is either horizontally or vertically polarized. The detection of each of these polarizations is done using a polarizing beam splitter ($\text{PBS}_1$) and two single-photon counting modules, D0 and D1, connected to multi-mode fibers (MMF) . The heralded photon is detected on Bob's side using $\text{PBS}_4$ and a two MMFs coupled to single-photon counting modules D2 or D3. 

With our setup, there are various possibilities to register a joint count between Alice and Bob: when $\mathrm{HWP}_A$ and $\mathrm{HWP}_B$ are set at 0$^{\circ}$ or at 22.5$^{\circ}$, there are joint counts only between D1 and D2 ( referred to as $D_{12}$) or joint counts between D0 and D3 (referred to as $D_{03}$). On the other hand, when $\mathrm{HWP}_A$ and $\mathrm{HWP}_B$ are set at different angles, D1 can have a joint count with either D2 or D3 (referred to as $D_{13}$) and similarly D0 can have a joint count with either D2 (referred to as $D_{02}$) or D3. These various possibilities constitute different $G^{2}(\tau)$ functions, shown in Appendix 3. All of them are measured by sending the output pulses from the detectors to a Time-to-digital converter (TDC), QuTools QuTAU, with a temporal resolution of 81~ps. From these measurements, one can obtain the values for $\tau_{0}$ and $\sigma$ that allow to recognize the HSP.

Once the criteria to recognize a HSP is established, it is possible to implement the BB84 protocol. This is done as follows: Alice and Bob randomly choose a wave plate position between 0$^{\circ}$ and 22.5$^{\circ}$. After the position in the wave plates is set, Alice and Bob register the detector that does click and the corresponding time stamping. When the time stamping matches, it indicates that there is a joint count, i.e., there is a HSP that can be used to generate the key. The bits for the key are assigned as follows: in Alice's arm, logical 0 and logical 1 are associated to clicks in D0 and D1, respectively. In Bob's arm, logical 0 and logical 1 are associated to clicks in D3 and D2, respectively. Further details on the data analysis are presented in Appendix~\ref{sec: appendix3}.
For each combination of positions of $HWP_A$ and $HWP_B$, there are various HSP. In our experiment we have an average of 9 HSP for each wave plate position. Each HSP leads to a bit. For the data reported here, we considered all those bits in the keys. This does not constitute a variation of the standard BB84 protocol, but makes our proof-of-principle demonstration more efficient. 

With the procedure described above, Alice and Bob have the information about the position of each wave-plate and the registered bits. This allows them to follow the steps of the ideal BB84 protocol and distribute a key. In our experimental proof-of-principle demonstration, we generated 5 keys of $\approx$ 1000 bits each. The average $\mathrm{QBER}$ value is 3.9 $\pm$ 0.3 $\%$. This value determines the minimum $\mathrm{QBER}$ value in our experiment. The fact that our experimental $\mathrm{QBER}$ is not zero is due to dark counts in the detectors, polarization imperfections and background noise. 
 
\begin{figure*}[t]
    \centering
    \includegraphics[width=1.9\columnwidth]{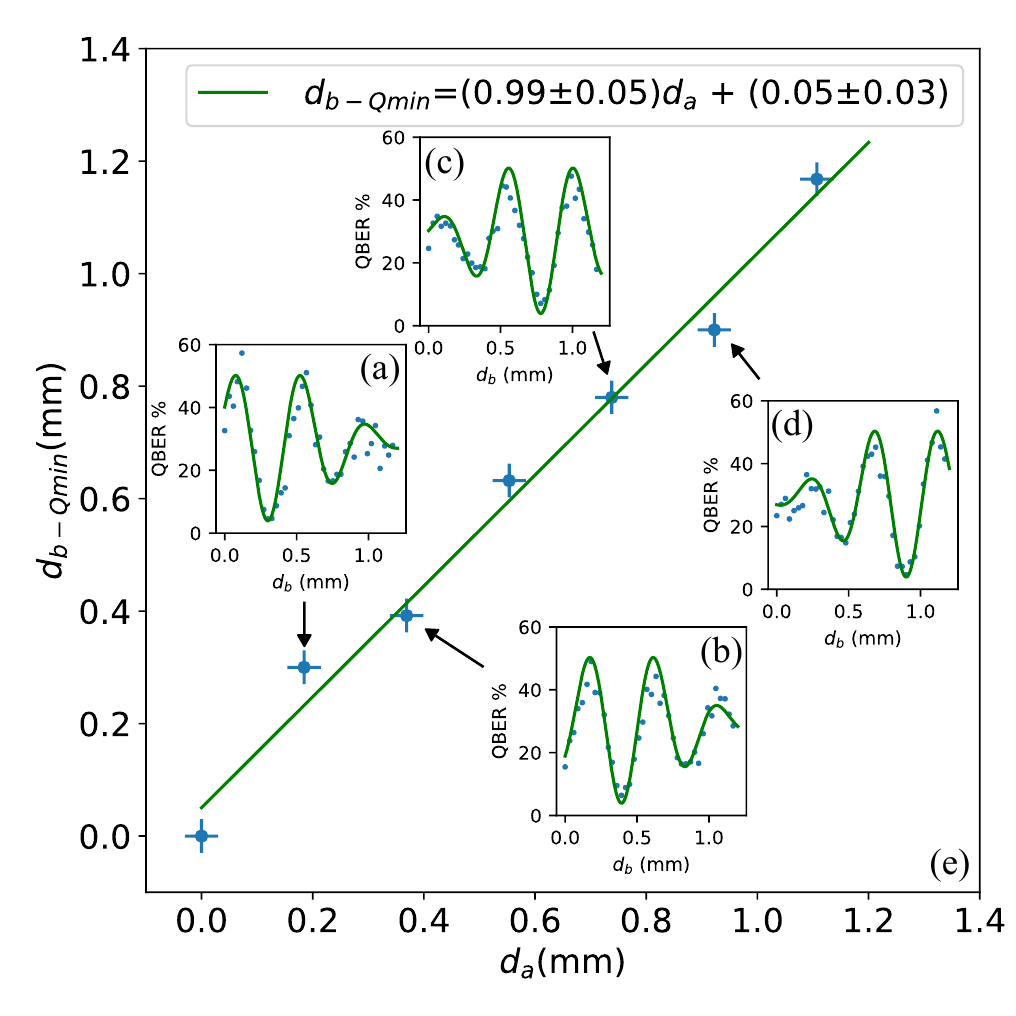}
    \caption{$\mathrm{QBER}$ in the controllable decoherence assisted scheme. In (a-d) the $\mathrm{QBER}$ is plotted, for different values of $d_a$, as a function of $d_b$. Each data point in every figure was made by averaging the QBER of 1000 bits. In (e), the position of the value of $d_b$ that minimizes the QBER is plotted against $d_a$.}
    \label{fig:QBER}
\end{figure*}

\subsection{Experiment 2: BB84 protocol under the controllable decoherence assisted scheme}
The second proof-of-principle experiment we present consists of introducing the controllable decoherence assisted scheme into the BB84 protocol. In order to do so, two dephasing channels are introduced to the setup by placing PBS2 and PBS3 as shown in the yellow box of Fig.~\ref{fig:setup}. 

Each dephasing channel is implemented using a polarizing tunable beam displacer (P-TBD) \cite{doi:10.1063/1.4914834}. This device 
takes an input polarized beam and divides it into two parallel beams with orthogonal polarizations. The P-TBD consists of two mirrors and a PBS mounted on a rotating platform. The angle that the platform is rotated determines the distance, $d$, that the two parallel beams are separated. The platform can rotate clockwise or counter-clockwise and this results in the two parallel beams being separated or getting closer, respectively. The P-TBD has been demonstrated to work as a controllable dephasing channel by coupling polarization and transverse momentum variables of light~\cite{Urrego:18}. The controllable feature comes from the fact that the P-TBD is a tunable device in which the distance $d$ acts as the control parameter.

On Alice's side, the P-TBD-A introduces controlled decoherence, governed by the parameter $d_a$, after the polarization state is prepared by $\text{HWP}_A$. On Bob's side, P-TBD-B is controlled by the parameter $d_b$ and introduces decoherence by rotating the platform in the opposite direction of P-TBD-A. With these two devices, the operators $\hat{U}(-d_b)$ and $\hat{U}(d_a)$ that appear in Eq.~\eqref{eq:operators} are implemented. The addition of P-TBDs to the experiment introduces an additional optical path, requiring a new temporal characterization of the coincidences via $G^{(2)}(\tau)$ (cf. Appendix 3).


To show that our controllable decoherence assisted scheme can be implemented, it is necessary to corroborate the validity of Eq.~\eqref{eq:QBER_scheme}. In our experimental implementation, we achieved a spatial profile 
$f(y)=(2/\pi \text{w}^{2})^{1/4}\exp(-y^{2}/\text{w}^{2})\exp(iq_{0}y)$. In the experimental implementation, a tilt in the propagation of the two orthogonally polarized beams introduces a transverse wavenumber $q_0$, which can be estimated from experimental data. Physically, $q_{0}$ accounts for the fact that the light beams do not impinge on the PBS of the dephasing channel perpendicularly. Taking into account this, 
Eq.~\eqref{eq:QBER_scheme} becomes
\begin{multline}
\mathrm{QBER}(d_{a},d_{b})=\\
\frac{1}{4}\left\{1-\exp\left[-2\frac{(d_{a}-d_{b})^{2}}{\text{w}^{2}}\right]\,\cos[2q_{0}(d_{a}-d_{b})]\right\}.
\label{new_QBER_scheme}
\end{multline}
The $\mathrm{QBER}$ values are measured for keys generated under the effect of different values of the parameters $d_a$ and $d_b$. Specifically, we measure the $\mathrm{QBER}$ for a setup in which we fix the value of $d_a$ and we scan the value of $d_b$. We repeat this measurement for seven different values of $d_{a}$. Four of them are shown in Fig.~\ref{fig:QBER}(a-d). The dots are the experimental data and the solid line corresponds to the theoretical curve according to Eq.~\eqref{new_QBER_scheme} using a value of $q_{0}$ = 6.87 $\pm$ 0.08 mm$^{-1}$. This value of $q_{0}$ corresponds to the average of seven values of $q_{0}$, each one obtained from fitting the experimental data to Eq.~\eqref{new_QBER_scheme}. 

From Fig~\ref{fig:QBER}(a-d), it is clear that the $\mathrm{QBER}$ has an oscillatory behaviour. For some values of $d_a$ and $d_b$ the $\mathrm{QBER}$ is higher that the one we found for the ideal BB84 protocol, this can be understood due to the presence of decoherence. Interestingly, as we expected for our controllable decoherence scheme, when $d_a= d_b$ low values of $\mathrm{QBER}$ are recovered. This is clearly demonstrated by Fig~\ref{fig:QBER}(e) where the value of $d_b$, when the $\mathrm{QBER}$ is minimum ($d_{b-Qmin }$), is plotted against the value of $d_a$. A straight line is clearly recognized demonstrating that indeed the $\mathrm{QBER}$ is minimum when $d_{a} = d_{b}$. It is worth mentioning that the minimum value of the $\mathrm{QBER}$ obtained using this protocol is always below 11$\%$, that guarantees that a secure and secret key can be distilled from the raw key after appropriate quantum error correction and private amplification \cite{Scarani2009}, providing unconditional security.

\section{conclusions\label{conclusions}}

We have presented a method that allows to reduce the amount of information that an eavesdropper can obtain in the BB84 protocol. This method is based on the introduction of decoherence in a controlled way using two dephasing channels.\\

We test the theoretical efficacy of this method by using the entangling probe attack and demonstrate that the R\'enyi information that Eve can obtain under the entangling probe attack is reduced for values of the QBER below the security limit of $\mathrm{QBER}$$<11\%$.\\

To illustrate the working principle of the controllable decoherence scheme, we have presented proof-of-principle demonstrations of the BB84 protocol using heralded single photons without and with a decoherence assisted scheme. In the first case we obtained $\mathrm{QBER}$= 3.9 $\pm$ 0.3 $\%$ averaging five keys of 1000 bits. In the second experiment, the controllable decoherence assisted scheme is used in the BB84 protocol and we observed that regardless of the presence of decoherence, it is possible to recover low $\mathrm{QBER}$ values.

\bibliographystyle{apsrev4-1} 
\bibliography{apssamp}

\begin{acknowledgments}

D.S and A.V acknowledge financial support from the Proyecto Semilla of the Facultad de Ciencias at the Universidad de los Andes, with identification codes INV-2022-142-2435 and INV-2022-143-2490. J.R.A. acknowledges funding by the European Union Horizon 2020 (Marie Sklodowska-Curie 765075-LIMQUET), PHOQUSING project GA no. 899544), and from the Plan France 2030 through the project ANR-22-PETQ-0006. This work is part of the R$\&$D project CEX2019-000910-S, funded by the Ministry of Science and innovation (MCIN/ AEI/10.13039/501100011033/). It has also been funded by Fundació Cellex, Fundació Mir-Puig, and from Generalitat de Catalunya through the CERCA program. We acknowledge support from the project QUISPAMOL (PID2020-112670GB-I00) funded by MCIN/AEI /10.13039/501100011033.
\end{acknowledgments}

\onecolumngrid

\newpage
\appendix

\subsection{Eve's Density matrix}

\label{sec:appendix1} In this appendix, we derive the explicit
form of the quantum state of Eve's probe photon when the controllable
decoherence-assisted scheme is used in the entangling probe attack.
When Alice sends an horizontal photon and Bob detects a photon with
the same polarization, the quantum state of Eve's photon is 
\begin{equation}
\hat{\rho}_{H}^{{\cal \left({\cal E}\right)}}=\mathrm{Tr}_{\mathrm{env}}\big\{\ket{\tilde{\psi}_{H}}_{{\cal B},{\cal E}}\bra{\tilde{\psi}_{H}}_{{\cal B},{\cal E}}\big\}
\end{equation}
where {[}see Eq.~(\ref{eq: CNOT_transformation_dec}(a){]}
\begin{equation}
\ket{\tilde{\psi}_{H}}_{{\cal B},{\cal E}}=\int dyf(y)\ket{y}_{{\cal B}}\Big(\sqrt{1-S^{2}}\ket{+}_{{\cal E}}-\frac{S}{\sqrt{2}}\ket{-}_{{\cal E}}\Big).
\end{equation}
One obtains 
\begin{align}
\hat{\rho}_{H}^{\left({\cal E}\right)} & =\frac{1}{1-S^{2}/2}\Bigg[(1-S^{2})\ket{+}_{{\cal {E}}}\bra{+}_{{\cal {E}}}+\frac{S^{2}}{2}\ket{-}_{{\cal {E}}}\bra{-}_{{\cal {E}}}\\
 & \hspace{2cm}-\frac{S\sqrt{1-S^{2}}}{\sqrt{2}}\ket{-}_{{\cal {E}}}\bra{+}_{{\cal {E}}}-\frac{S\sqrt{1-S^{2}}}{\sqrt{2}}\ket{+}_{{\cal {E}}}\bra{-}_{{\cal {E}}}\Bigg].\nonumber 
\end{align}
Similarly, for a photon with vertical polarization, one has $\hat{{\rho}}_{V}^{{(\cal {E}})}=\text{Tr}_{\text{env}}\big\{\ket{\tilde{\psi}_{V}}_{{\cal B},{\cal E}}\bra{\tilde{\psi}_{V}}_{{\cal B},{\cal E}}\big\}$
with {[}see Eq.~(\ref{eq: CNOT_transformation_dec}(b){]} 
\begin{equation}
\ket{\tilde{\psi}_{V}}_{{\cal {\cal B}},{\cal E}}=\int dyf(y)\ket{y}_{{\cal B}}\Big(\sqrt{1-S^{2}}\ket{+}_{{\cal E}}+\frac{S}{\sqrt{2}}\ket{-}_{{\cal E}}\Big)
\end{equation}
that yields 
\begin{align}
\hat{\rho}_{V}^{{\cal \left({\cal E}\right)}} & =\frac{1}{1-S^{2}/2}\Bigg[(1-S^{2})\ket{+}_{{\cal {E}}}\bra{+}_{{\cal {E}}}+\frac{S^{2}}{2}\ket{-}_{{\cal {E}}}\bra{-}_{{\cal {E}}}\\
 & \hspace{2cm}+\frac{S\sqrt{1-S^{2}}}{\sqrt{2}}\ket{-}_{{\cal {E}}}\bra{+}_{{\cal {E}}}+\frac{S\sqrt{1-S^{2}}}{\sqrt{2}}\ket{+}_{{\cal {E}}}\bra{-}_{{\cal {E}}}\Bigg]\nonumber 
\end{align}
For diagonal polarization, $\hat{\rho}_{D}^{{\cal \left({\cal E}\right)}}=\text{Tr}_{\text{env}}\big\{\ket{\tilde{\psi}_{D}}_{{\cal {\cal B}},{\cal E}}\bra{\tilde{\psi}_{D}}_{{\cal B},{\cal E}}\big\}$
with {[}see Eq.~(\ref{eq: CNOT_transformation_dec}(c){]} 
\begin{equation}
\ket{\tilde{\psi}_{D}}_{{\cal B},{\cal E}}=\frac{1}{2}\int dy\Bigg[f(y)\Big(2\sqrt{1-S^{2}}\ket{+}_{{\cal E}}\Big)+\frac{S}{\sqrt{2}}\Big(f(y-2d)+f(y+2d)\Big)\ket{-}_{{\cal E}}\Bigg]\ket{y}_{{\cal B}}
\end{equation}
The quantum state is now 
\begin{align}
\hat{\rho}_{D}^{{\cal \left({\cal E}\right)}} & =\frac{1}{1+(\gamma_{1}/8-1)S^{2}}\Bigg[(1-S^{2})\ket{+}_{{\cal {E}}}\bra{+}_{{\cal {E}}}+\frac{S^{2}\gamma_{1}}{8}\ket{-}_{{\cal {E}}}\bra{-}_{{\cal {E}}}\\
 & \hspace{2cm}+\frac{S\sqrt{1-S^{2}}}{2\sqrt{2}}\gamma_{2}\ket{-}_{{\cal {E}}}\bra{+}_{{\cal {E}}}+\frac{S\sqrt{1-S^{2}}}{2\sqrt{2}}\gamma_{2}^{*}\ket{+}_{{\cal {E}}}\bra{-}_{{\cal {E}}}\Bigg],\nonumber 
\end{align}
where 
\begin{equation}
\gamma_{1}=\int_{-\infty}^{\infty}dy|f(y+2d)+f(y-2d)|^{2}
\end{equation}
and 
\begin{equation}
\gamma_{2}=\int_{-\infty}^{\infty}dyf^{*}(y)\Big[f(y+2d)+f(y-2d)\Big]
\end{equation}
Finally, for anti-diagonal polarization, $\hat{\rho}_{A}^{{\cal \left({\cal E}\right)}}=\text{Tr}_{\text{env}}\big\{\ket{\tilde{\psi}_{A}}_{{\cal B},{\cal E}}\bra{\tilde{\psi}_{A}}_{{\cal B},{\cal E}}\big\}$
with {[}see Eq.~(\ref{eq: CNOT_transformation_dec}(d){]} 
\begin{equation}
\ket{\tilde{\psi}_{A}}_{{\cal {\cal B}},{\cal E}}=\frac{1}{2}\int dy\Bigg[f(y)\Big(\ket{T_{-}}_{{\cal E}}+\ket{T_{+}}_{{\cal E}}\Big)-\Big(f(y-2d)+f(y+2d)\Big)\ket{T_{E}}_{{\cal E}}\Bigg]\ket{y}_{{\cal B}}
\end{equation}
The quantum state of Eve's photon is 
\begin{align}
\hat{\rho}_{A}^{{\cal \left({\cal E}\right)}} & =\frac{1}{1+(\gamma_{1}/8-1)S^{2}}\Bigg[(1-S^{2})\ket{+}_{{\cal {E}}}\bra{+}_{{\cal {E}}}+\frac{S^{2}\gamma_{1}}{8}\ket{-}_{{\cal {E}}}\bra{-}_{{\cal {E}}}\\
 & \hspace{2cm}-\frac{S\sqrt{1-S^{2}}}{2\sqrt{2}}\gamma_{2}\ket{-}_{{\cal {E}}}\bra{+}_{{\cal {E}}}-\frac{S\sqrt{1-S^{2}}}{2\sqrt{2}}\gamma_{2}^{*}\ket{+}_{{\cal {E}}}\bra{-}_{{\cal {E}}}\Bigg]\nonumber 
\end{align}

\subsection{Calculation of the trace distances}

\label{sec:appendix2} In order to calculate the Rényi information,
it is necessary to obtain the trace distances $D(\rho_{H}^{{\cal \left({\cal E}\right)}},\rho_{V}^{{\cal \left({\cal E}\right)}})$
and $D(\rho_{D}^{{\cal \left({\cal E}\right)}},\rho_{A}^{{\cal \left({\cal E}\right)}})$,
which corresponds to the quantum states that Eve needs to discriminate. The trace distance can be calculated by \begin{equation}
D(\hat{\rho}_{1}^{{\cal \left({\cal E}\right)}},\hat{\rho}_{2}^{{\cal \left({\cal E}\right)}})=\frac{1}{2}\sum_{i}^{n}|\lambda^{(1,2)}_{i}|,
\end{equation}
where $\lambda_{i}^{(1,2)}$ are the eigenvalues of $\hat{\rho}_{(1,2)}^{{\cal \left({\cal E}\right)}}=\hat{\rho}_{2}^{{\cal \left({\cal E}\right)}}-\hat{\rho}_{1}^{{\cal \left({\cal E}\right)}}$.

The eigenvalues of the density matrix 
\begin{align}
\hat{\rho}_{HV}^{{\cal \left({\cal E}\right)}}=\hat{\rho}_{V}^{{\cal \left({\cal E}\right)}}-\hat{\rho}_{H}^{{\cal \left({\cal E}\right)}}=\frac{2}{1-S^{2}/2}\Bigg[\frac{S\sqrt{1-S^{2}}}{\sqrt{2}}\ket{-}_{{\cal {E}}}\bra{+}_{{\cal {E}}}+\frac{S\sqrt{1-S^{2}}}{\sqrt{2}}\ket{+}_{{\cal {E}}}\bra{-}_{{\cal {E}}}\Bigg],
\end{align}
are $\lambda_{1,2}^{(H,V)}=\pm2\sqrt{2}\,S\,\sqrt{1-S^{2}}/(2-S^{2})$, so 
\begin{equation}
D(\hat{\rho}_{H}^{{\cal \left({\cal E}\right)}},\hat{\rho}_{V}^{{\cal \left({\cal E}\right)}})=2\sqrt{2}\frac{S\sqrt{1-S^{2}}}{2-S^{2}}
\end{equation}
The eigenvalues of the density matrix 
\begin{align}
\hat{\rho}_{DA}^{{\cal \left({\cal E}\right)}}=\hat{\rho}_{D}^{{\cal \left({\cal E}\right)}}-\hat{\rho}_{A}^{{\cal \left({\cal E}\right)}}=\frac{1}{1+(\gamma_{1}/8-1)S^{2}}\Bigg[\frac{S\sqrt{1-S^{2}}}{\sqrt{2}}\gamma_{2}^{*}\ket{-}_{{\cal {E}}}\bra{+}_{{\cal {E}}}+\frac{S\sqrt{1-S^{2}}}{\sqrt{2}}\gamma_{2}\ket{+}_{{\cal {E}}}\bra{-}_{{\cal {E}}}\Bigg],
\end{align}
are $\lambda_{1,2}^{(D,A)}=\pm4\sqrt{2}\,S\,\sqrt{1-S^{2}}\,\gamma_{2}/[8+(\gamma_{1}-8)S^{2}]$, so
\begin{equation}
D(\hat{\rho}_{D}^{\left({\cal E}\right)},\hat{\rho}_{A}^{{\cal \left({\cal E}\right)}})=4\sqrt{2}\frac{S\sqrt{1-S^{2}}}{8+(\gamma_{1}-8)S^{2}}\gamma_{2}.
\end{equation}
For the case of a function $f(y)$ with spatial Gaussian shape, 
\begin{equation}
f(y)=\Big(\frac{2}{\pi\text{w}^{2}}\Big)^{1/4}\exp\big[-y^{2}/\text{w}^{2}\big],
\end{equation}
the parameters $\gamma_{1}$ and $\gamma_{2}$ become $\gamma_{1}=2+2\gamma_{0}^{4}$
and $\gamma_{2}=2\gamma_{0}$ with $\gamma_{0}=\exp(-2d^{2}/\text{w}^{2})$. 

\subsection{Obtaining the key from time stampings}\label{sec: appendix3}

\begin{figure}[H]
    \centering
    \includegraphics[scale = 0.4]{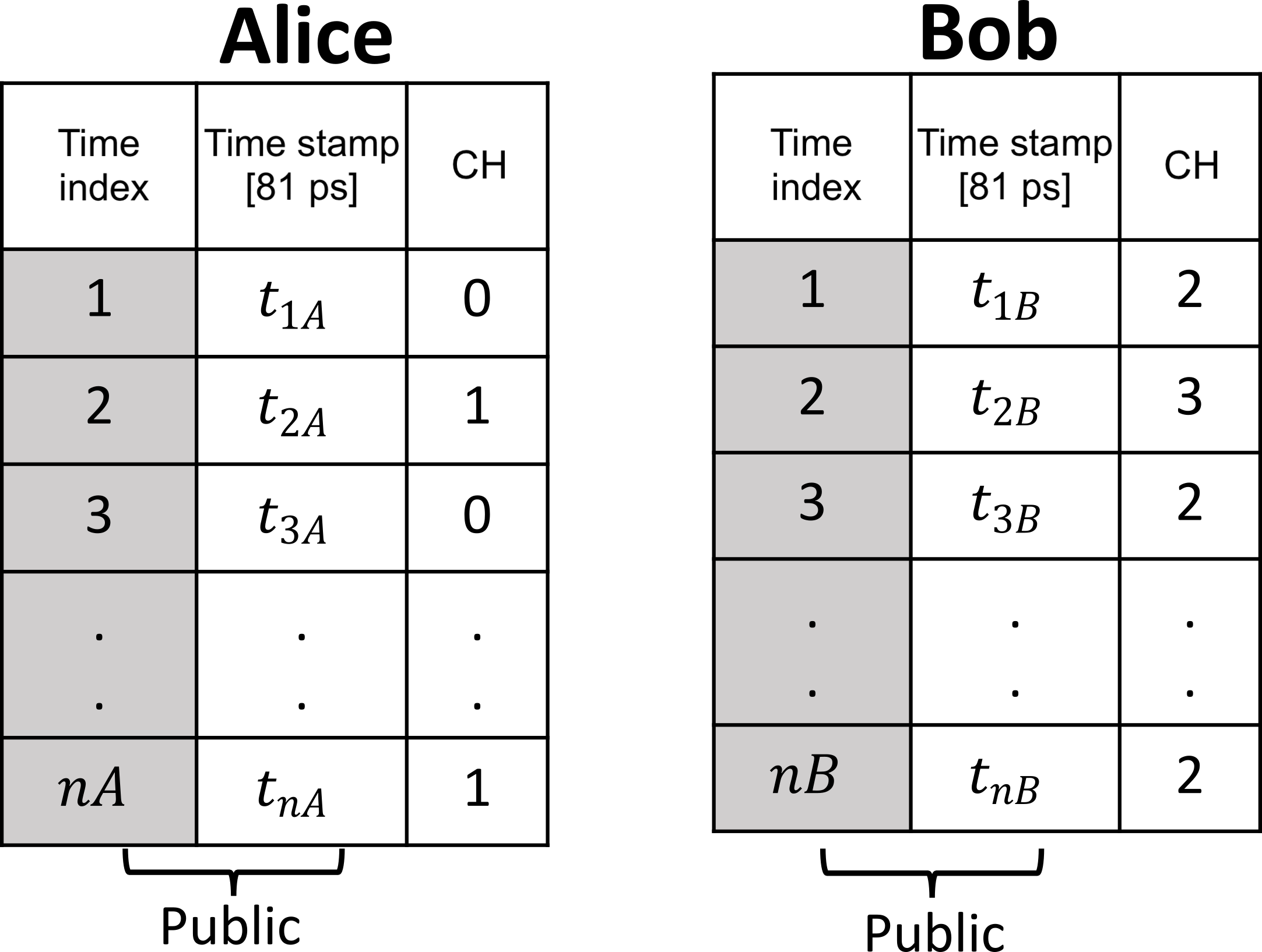}
    \caption{Scheme of data analysis to recognize HSPs. Alice and Bob add one index to each event. Afterwards, they share publicly a list with the time index and the time of each click.}
    \label{fig:Time-index}
\end{figure}

In this appendix, we explain the detailed process of obtaining the key from the time stamping list. The process is as follows: after the position in the wave plates is set, Alice and Bob generate a file that contains the position of its own wave plate and a list that has the time stampings and the detector that produces the click. Afterwards, computationally Alice and Bob add one column to its own list that contains a number that indexes the position of each element of the list, this is illustrated by the gray column in Fig.~\ref{fig:Time-index}.

\begin{figure}[H]
    \centering
    \includegraphics[scale = 0.5]{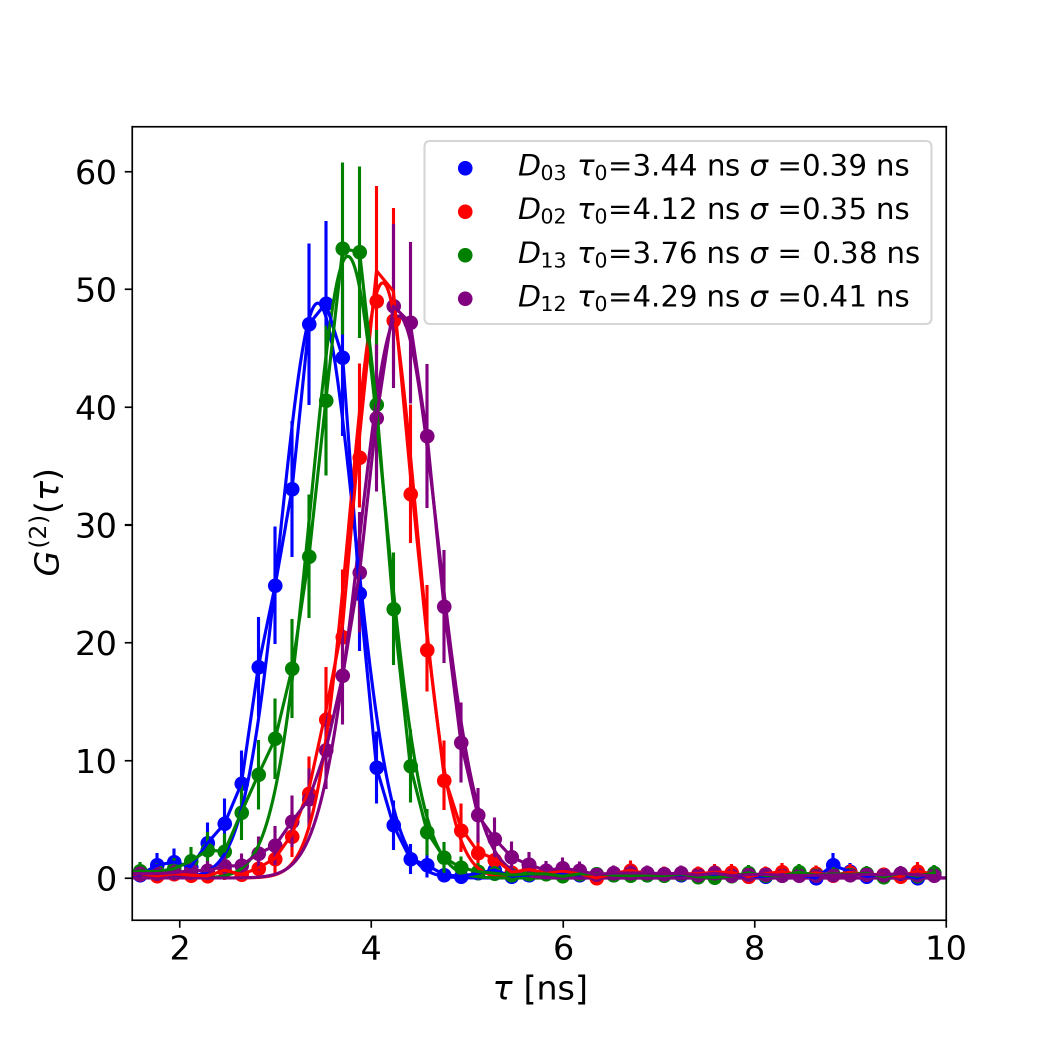}
    \caption{Temporal characterization used to recognize heralded single-photons in the standard BB84 protocol.}
    \label{fig:G2}
\end{figure}

\begin{figure}[H]
    \centering
    \includegraphics[scale = 0.5]{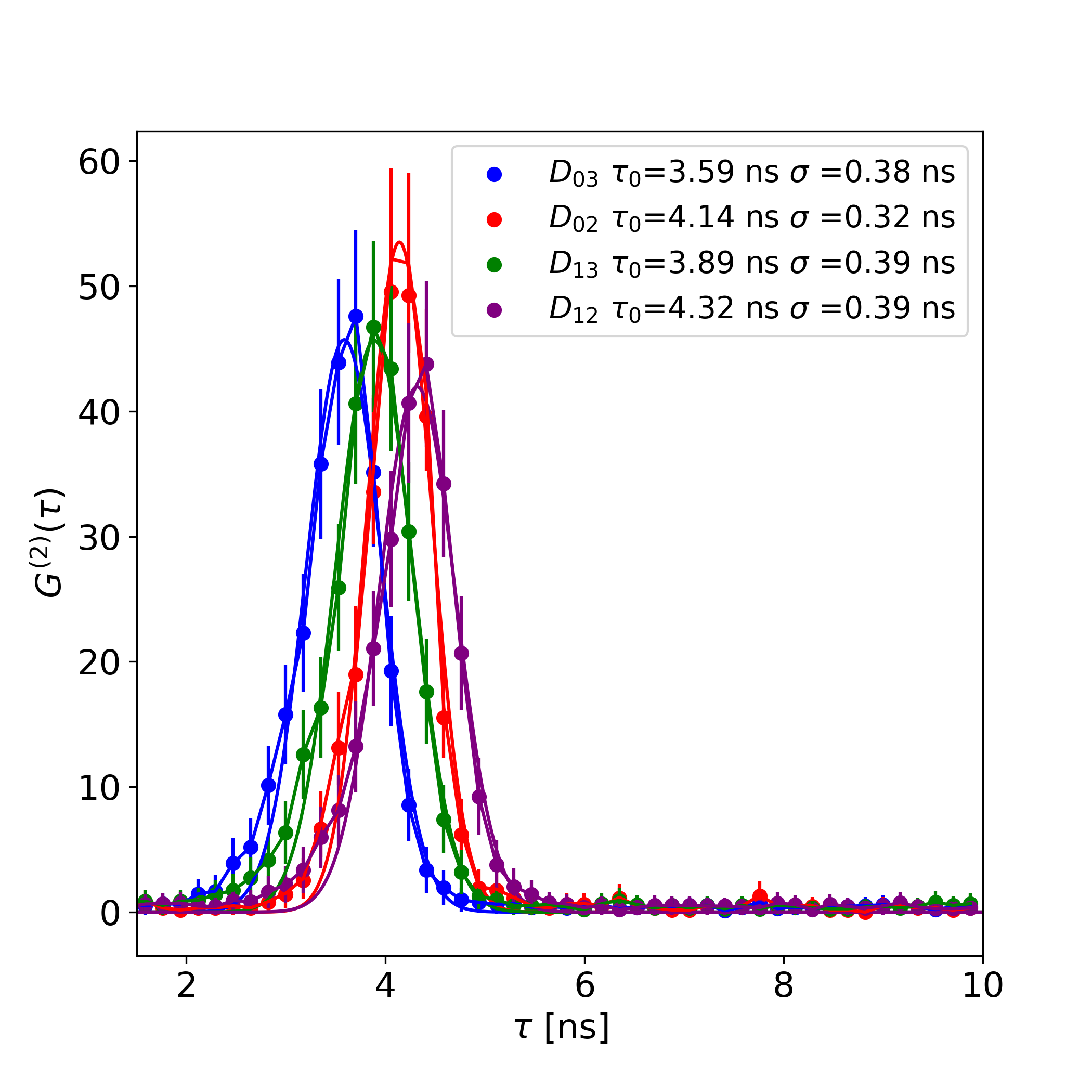}
    \caption{Temporal characterization used to recognize heralded single-photons when the controllable decoherence-assisted scheme is used.}
    \label{fig:G2_DEC}
\end{figure}

To identify HSPs, Alice and Bob make public the portion of their own list that contains time stamps and time indexes. When the standard BB84 protocol is implemented, a HSP is identified as a joint count among $\tau_{0} \pm 2 \sigma$ in any of the $G^{(2)}$ measurements of  Fig.~\ref{fig:G2}. On the other hand, when the P-TBDs are introduced in the controllable decoherence-assisted scheme, the recognition of a HSP is given by any joint count among $\tau_{0} \pm 2 \sigma$ in any of the $G^{(2)}$ measurements of Fig.~\ref{fig:G2_DEC}.

\begin{figure}[H]
    \centering
    \includegraphics[scale = 0.4]{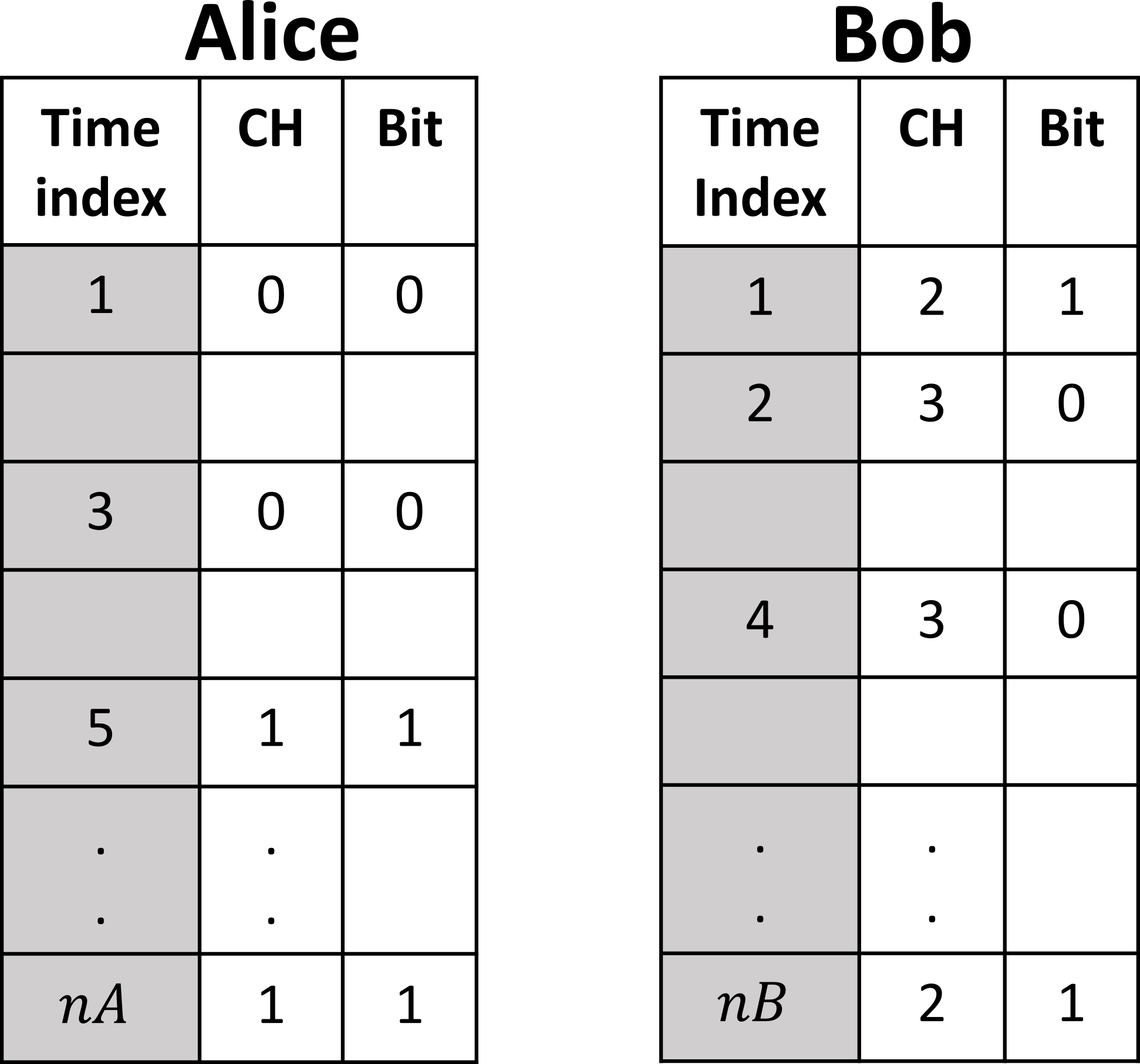}
    \caption{ Scheme of data analysis to generate the shared key. The empty boxes are due to the fact that the event was not taken into account because it is not a joint count.}
    \label{fig:A_B_list}
\end{figure}

After the identification of HSPs, Alice and Bob save the time indexes of the joint counts without revealing the detector. Once the time indexes are saved, Alice and Bob assign bits to the detectors that led to a joint count: in Alice's arm, logical 0 and logical 1 are associated to clicks in D0 and D1, respectively. In Bob's arm, logical 0 and logical 1 are associated to clicks in D3 and D2, respectively. This is illustrated in Fig~\ref{fig:A_B_list}. The bits assigned will constitute the key.

\end{document}